\documentclass[aps,twocolumn,raggedbottom,prb,showpacs,nobalancelastpage,amssymb,superscriptaddress]{revtex4}
\usepackage{amsmath}
\usepackage{amssymb}
\usepackage{natbib}
\usepackage{graphicx}
\usepackage{epstopdf}
\usepackage{color}
\usepackage{times}
\usepackage[colorlinks,bookmarks=false,citecolor=blue,linkcolor=red,urlcolor=blue]{hyperref}

\def\Jr{J_\mathrm{rung}}
\def\Jl{J_\mathrm{leg}}

\begin{document}
\title{Entanglement spectra of coupled $S=1/2$ spin chains in a ladder geometry}
\author{Andreas M. L\"auchli}
\email{andreas.laeuchli@uibk.ac.at}
\affiliation{Institut f\"ur Theoretische Physik, 
Universit\"at Innsbruck, A-6020 Innsbruck, Austria}
\affiliation{Max-Planck-Institut f\"ur Physik komplexer Systeme, 
N\"othnitzer Str.\ 38, D-01187 Dresden, Germany}
\author{John Schliemann}
\email{john.schliemann@physik.uni-regensburg.de} 
\affiliation{Institute for Theoretical Physics, University of Regensburg, D
-93040 Regensburg, Germany}
\date{\today}

\begin{abstract} 
We study the entanglement spectrum of spin-$1/2$ XXZ ladders both analytically
and numerically. Our analytical approach is based on perturbation theory
starting either from the limit of strong rung coupling, or from the opposite
case of dominant coupling along the legs. In the former case we find to 
leading order that the entanglement Hamiltonian is also of nearest-neighbor XXZ form
although with an in general renormalized anisotropy. For the cases of
XX and isotropic Heisenberg ladders no such renormalization takes place. In the
Heisenberg case the second order correction to the entanglement Hamiltonian consists of
a renormalization of the nearest neighbor coupling plus an unfrustrated next nearest 
neighbor coupling. In the opposite regime of strong coupling along the legs, we point out an 
interesting connection of the entanglement spectrum to the Lehmann 
representation of single chain spectral functions of operators appearing in the 
physical Hamiltonian coupling the two chains.
\end{abstract}
\pacs{75.10.Jm, 03.67.-a} 
\maketitle

\section{Introduction}

In the last decade many-body physics has been substantially enriched
by the concept of entanglement, whose extensive and systematic study originated
in the field of quantum information theory \cite{Amico08}. In particular,
the notion of the {\em entanglement spectrum}\cite{Li08} 
has led to novel insights
in the physics of various many-body systems. These include
quantum Hall monolayers at fractional filling
\cite{Li08,Regnault09,Zozulya09,Lauchli10,Thomale10a,Sterdyniak10,Thomale10b,Chandran11,Sterdyniak11,Qi11}, 
quantum Hall bilayers at filling factor $\nu=1$\cite{Schliemann11}, 
spin systems of one
\cite{Calabrese08,Pollmann10a,Pollmann10b,Thomale09,Poilblanc10,Peschel11}
and two \cite{Yao10,Cirac11,Huang11} spatial dimensions,
and topological insulators \cite{Fidkowski10,Prodan10}. Other topics
recently covered encompass rotating Bose-Einstein condensates \cite{Liu11},
coupled Tomonaga-Luttinger liquids \cite{Furukawa11}, and systems
of Bose-Hubbard\cite{Deng11} and complex paired superfluids\cite{Dubail11},

In Ref.~\onlinecite{Poilblanc10} Poilblanc observed that chain-chain 
entanglement spectra in two-leg spin ladders are remarkably similar
to the {\em energy} spectrum of a single Heisenberg chain. Furthermore he 
found that the fitted effective inverse temperature depends 
on the ratio of the leg to the rung couplings and vanishes in the limit of 
strong rung coupling. 
In a parallel study of entanglement spectra of quantum hall bilayers at 
$\nu=1$ one of the present authors observed a similarly striking 
analogy between the entanglement spectrum of a single layer and the energy 
spectrum of a single physical layer at $\nu=1/2$.\cite{Schliemann11}

In this paper we study the entanglement spectrum of two coupled XXZ chains 
analytically in two limiting cases a) the case of strong rung coupling
and b) the case of weak rung coupling. In case a) we find that
the entanglement spectrum is described to leading order by an entanglement 
Hamiltonian of the XXZ form, albeit with an in general renormalized effective 
anisotropy (and thus independently reproducing the recently posted result of 
Ref.~\onlinecite{Peschel11}). For the particular cases of XX and Heisenberg ladders the 
anisotropy is unaltered, and we arrive at the conclusion that in these cases 
the entanglement Hamiltonian is indeed the physical Hamiltonian restricted
to a single chain. Moreover, we derive explicit results for the next-to-leading
order exhibiting deviations of the entanglement Hamiltonian from 
nearest-neighbors XXZ coupling. In the second case (b) we point out an interesting connection 
between the entanglement spectrum and the Lehmann representation of single
chain spectral functions of the operators contained in the physical Hamiltonian 
coupling the two chains.

We consider an XXZ spin ladder Hamiltonian ${\cal H}={\cal H}_0+{\cal H}_1$ with
\begin{eqnarray}
\label{eqn:Hamiltonian}
{\cal H}_0 & = & \Jr \sum_m\Bigl[\frac{1}{2}(S^+_{m,1} S^-_{m,2} + h.c.)
+\Delta S^z_{m,1}S^z_{m,2}\Bigr]\,,\\
{\cal H}_1 & = & \Jl \sum_{m,\nu}\Bigl[\frac{1}{2}(S^+_{m,\nu}  
S^-_{m+1,\nu} + h.c.)\nonumber
+\Delta S^z_{m,\nu}S^z_{m+1,\nu}\Bigr]
\end{eqnarray}
describing the coupling along the rungs and legs, respectively. The sites
are labeled by $(m,\nu)$, where $m\in\{1,\ldots,L\}$ denotes the position 
within chain $\nu = 1,2$.
All spin-$1/2$ operators are taken to be dimensionless such that 
$\Jr$, $\Jl$ have
dimension of energy, and $\Delta$ is a dimensionless XXZ anisotropy parameter.
In the following we will always assume a ladder of length $L$ with
periodic boundary conditions if not specified otherwise. In the following we limit our
discussion to the antiferromagnetic regime $\Jl,\Jr \ge 0$. Apart from the special case
$\Jl=1,\Jr=0$ (corresponding to two decoupled critical $S=1/2$ chains), the system is
gapped in this regime, with exponentially decaying correlations.

%%%%%%%%%%%%%%%%%
\begin{figure}
\centerline{\includegraphics[width=0.9\linewidth]{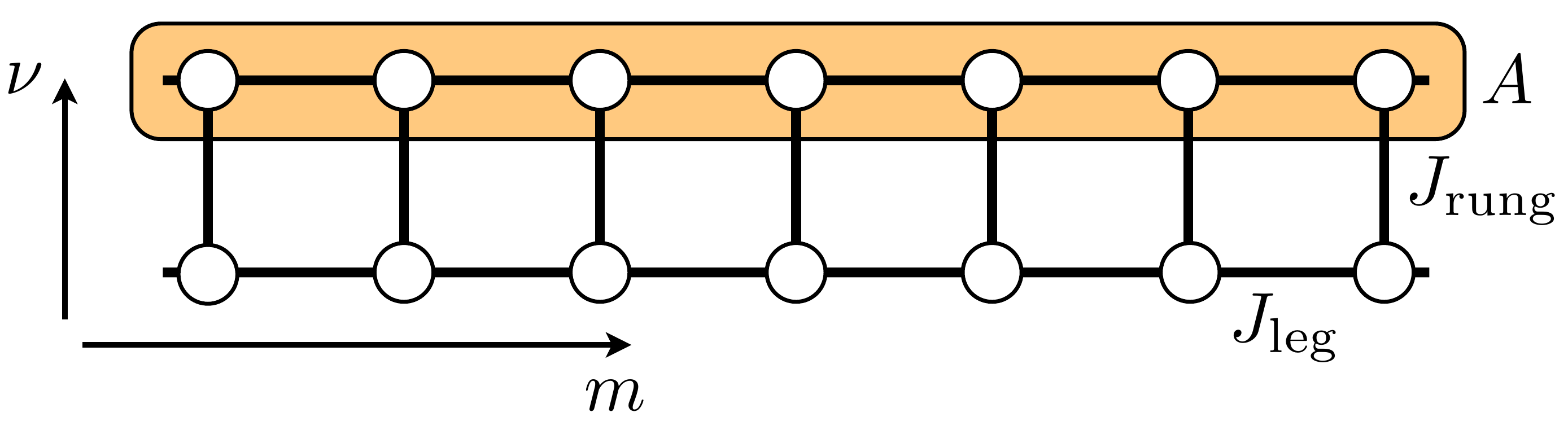}}
\caption{(Color online) Two leg spin ladder consider in this paper. The chains are labeled 
by $\nu=1,2$, while the runs are labeled by $m=1,\ldots L$. We consider the entanglement
spectrum in the illustrated chain-chain bipartition.}
\label{fig:ladder_setup}
\end{figure}
%%%%%%%%%%%%%%%%%

\section{Exact Diagonalization Results}

%%%%%%%%%%%%%%%%%
\begin{figure}
\centerline{\includegraphics[width=0.91\linewidth]{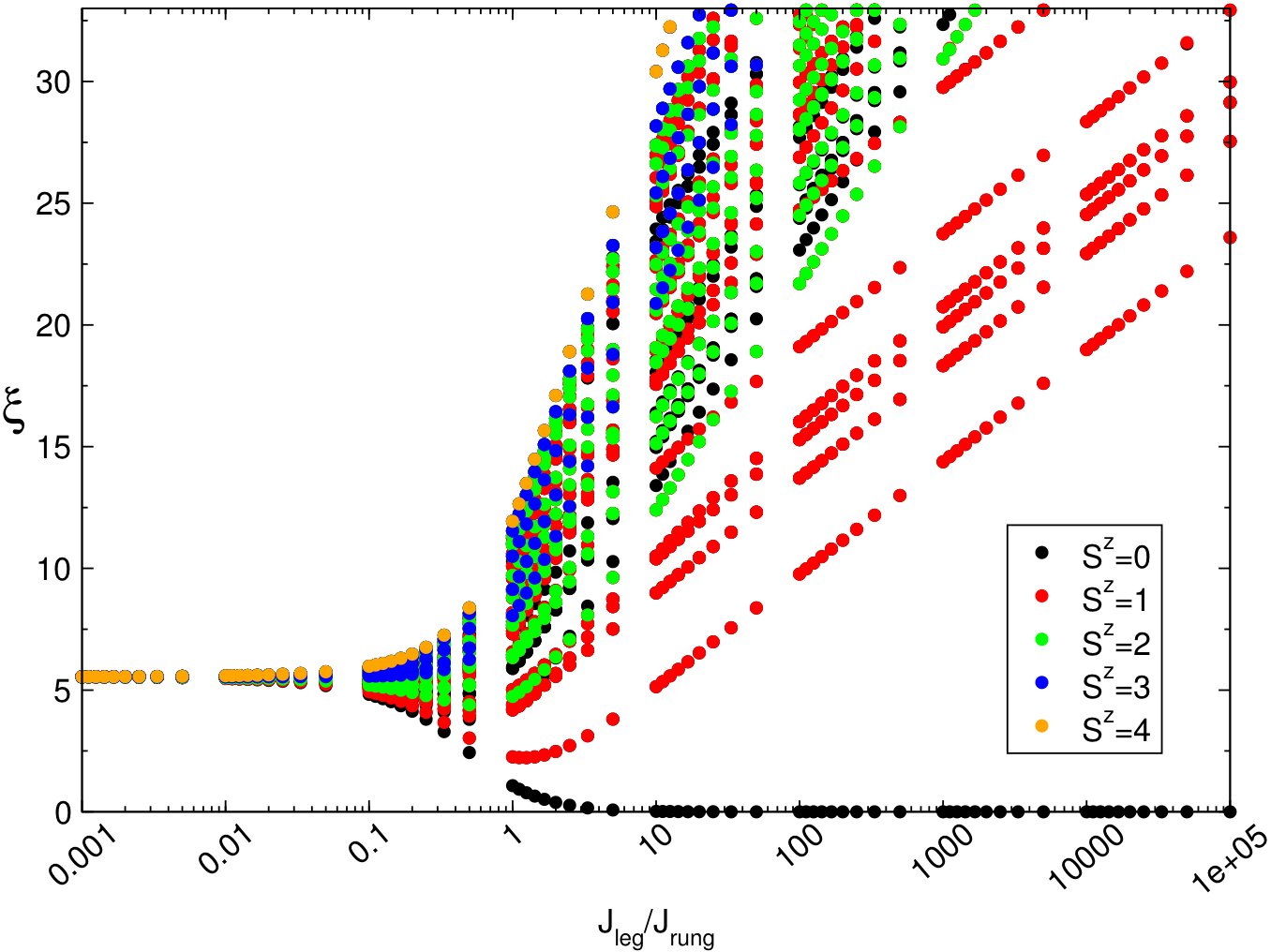}}
\vspace{0.3cm}
\centerline{\includegraphics[width=\linewidth]{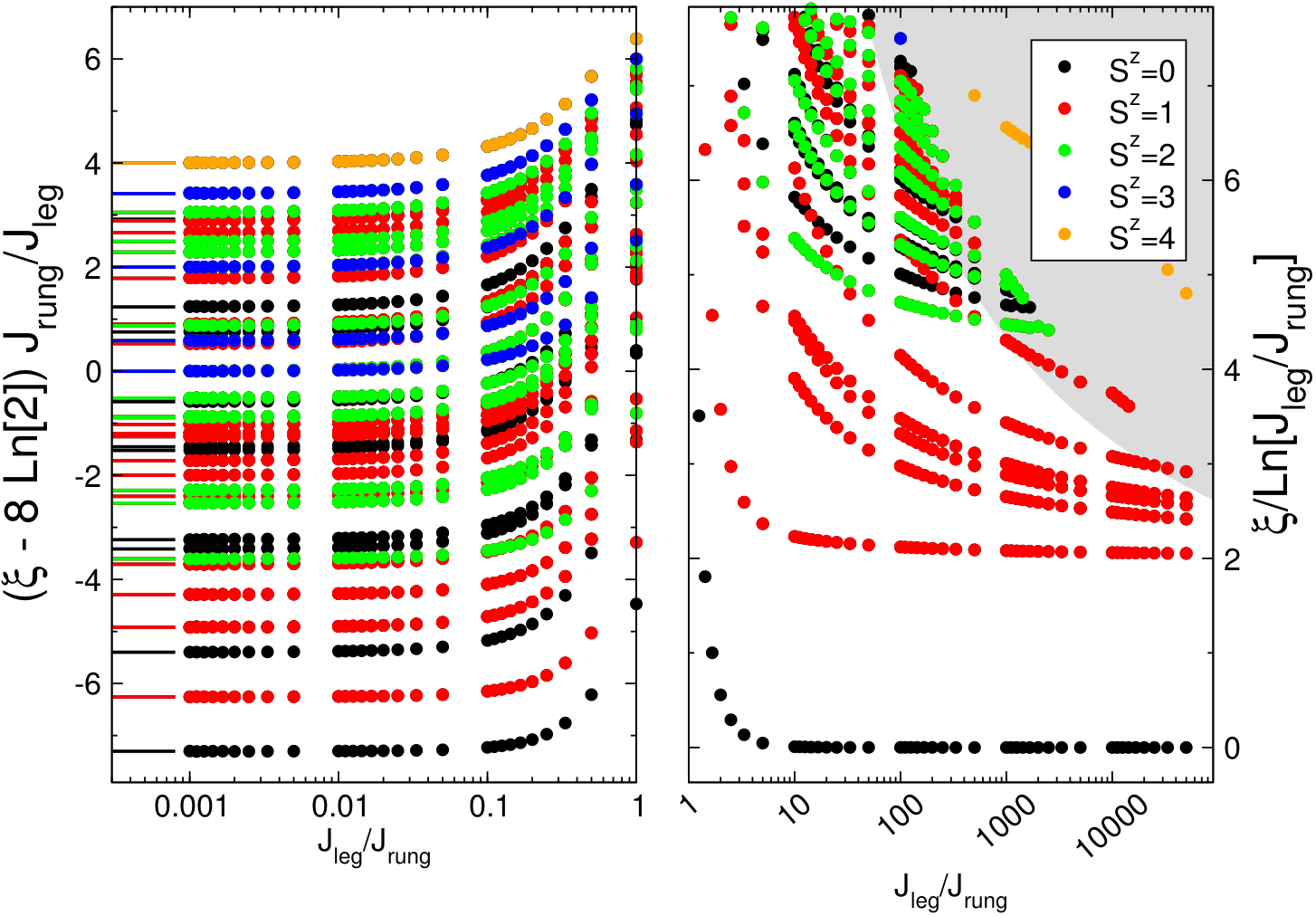}}
\caption{(Color online) Overview of the entanglement spectrum of the 
reduced density matrix of a single chain (cf. setup in 
Fig.~\ref{fig:ladder_setup}) 
in a Heisenberg ($\Delta=1$) spin ladder for different 
$S^z_\mathrm{tot}=S_\mathrm{tot}$ sectors as a function of $\Jl/\Jr$. 
The system size is $L=8$.
While in the upper panel the bare entanglement spectrum is shown, in the 
lower panels the leading asymptotic behavior in the two limiting cases
$\Jl/\Jr \rightarrow 0$ and $\Jr/\Jl\rightarrow 0$ is highlighted by 
appropriate shifts and rescaling indicated on the $y-$axes. The filled dots in all 
panels denote numerical entanglement spectrum levels, while the constant lines in the lower 
left panel highlight twice the eigenvalues of a single $L=8$ $S=1/2$ 
Heisenberg chain. The grey region in the lower right panel indicates the 
part of the rescaled entanglement spectrum affected by the finite (double)
precision arithmetic used in the numerical calculations. 
}
\label{fig:es_D1}
\end{figure}
%%%%%%%%%%%%%%%%%

Before discussing our analytical results in the two limiting cases we present an 
overview of the full $\Jl/\Jr$ dependence of the entanglement spectrum 
$\xi:=-\log(\lambda)$ obtained from the eigenvalues $\lambda$
of the reduced density matrix of a single chain in a Heisenberg 
($\Delta=1$) spin ladder (cf. bipartition setup shown in Fig.~\ref{fig:ladder_setup}). In the region $\Jl/\Jr < 1$ selected 
numerical spectra have already been presented in Refs.~\onlinecite{Poilblanc10,Cirac11}.

In the upper panel of Fig.~\ref{fig:es_D1} the bare entanglement spectrum for a $L=8$ ladder
is shown as a function of $\Jl/\Jr$ (note the logarithmic $x$-axis), where the levels are labeled 
according to their $S^z$ quantum number~\footnote{In the chain-chain bipartition setup the 
chain momentum is also a good quantum number, but we do not use this quantum number here 
for the sake of a simple presentation.}.
In the limit $\Jl/\Jr \rightarrow 0^+$ all the $2^L$ levels of the entanglement spectrum collapse
onto a single value $L \ln 2$, and for finite $\Jl/\Jr$ they start to spread. For $\Jl/\Jr \gtrsim
1$ the spectrum rearranges, and a notable feature is that the lowest levels above
the singlet ground states are a sizable set of $S=1$ triplets, displaying a common slope of $2$ when
plotted as a function of $\ln(\Jl/\Jr)$. Note that this is in contrast with the behavior at small $\Jl/\Jr$,
where there is only one triplet located below the second singlet in the entanglement spectrum.

In the lower panels of Fig.~\ref{fig:es_D1} we highlight the behavior in the two limit cases by appropriately
rescaling and shifting the entanglement spectra. In the lower left panel we also show the energy spectrum
of a $L=8$ Heisenberg chain (multiplied by two) to demonstrate the remarkable agreement between the
entanglement spectrum and the energy spectrum in this particular limit. As just discussed the structure of
the entanglement spectrum is somewhat different in the opposite limit, and the analogy with the energy spectrum
of a single chain apparently breaks down.

We now provide an analytical justification of the reported behavior based on perturbation theory around the two 
limits (a) and (b) described above.

\section{Strong rung coupling limit}
 Let us
now treat ${\cal H}_1$ as a perturbation to ${\cal H}_0$ with 
antiferromagnetic coupling, $\Jr>0$. The unperturbed
ground state reads
\begin{equation}
|0\rangle=\bigotimes_m |s_{m}\rangle
\end{equation}
using obvious notation for singlet and triplet states on the rungs,
\begin{eqnarray}
|s_m\rangle & = & \frac{1}{\sqrt{2}}
\left(\mid \uparrow\rangle_{m,1}\mid \downarrow\rangle_{m,2}
-\mid \downarrow\rangle_{m,1}\mid \uparrow\rangle_{m,2}\right)\,,\\
|t_m^+\rangle & = & \mid \uparrow\rangle_{m,1}\mid \uparrow\rangle_{m,2}\,,\\
|t_m^0\rangle & = & \frac{1}{\sqrt{2}}
\left(\mid \uparrow\rangle_{m,1}\mid \downarrow\rangle_{m,2}
+\mid \downarrow\rangle_{m,1}\mid \uparrow\rangle_{m,2}\right)\,,\\
|t_m^-\rangle & = & \mid \downarrow\rangle_{m,1}\mid \downarrow\rangle_{m,2}\,.
\end{eqnarray}

\subsection{First order}

The first-order correction to the ground state can be obtained by elementary
calculation,
\begin{eqnarray}
& & |1\rangle=+\frac{\Jl}{4 \Jr}\sum_m\Biggl[\frac{2}{1+\Delta}
\left(\cdots|t_m^+\rangle|t_{m+1}^-\rangle
\cdots\right)\nonumber\\
& & \qquad\qquad+\frac{2}{1+\Delta}
\left(\cdots|t_m^-\rangle|t_{m+1}^+\rangle
\cdots\right)\nonumber\\
& & \qquad\qquad-\Delta
\left(\cdots|t_m^0\rangle|t_{m+1}^0\rangle
\cdots\right)\Biggr]
\end{eqnarray}
where the dots denote singlet states on each rung not explicitly
specified. The reduced density operator is obtained by tracing out
one of the legs from $\rho=(|0\rangle+|1\rangle)(\langle 0|+\langle 1|)$ 
and is given within
first order in $\Jl/\Jr$ by
\begin{eqnarray}
& & \rho_{\rm red}^{(1)}=\frac{1}{2^L}\Biggl(1-\frac{4 \Jl}{\Jr(1+\Delta)}
\sum_m\Bigl[(S^+_{m}S^-_{m+1} + h.c.) \nonumber\\
& & \qquad\qquad\qquad+\frac{1}{2}(\Delta+\Delta^2)
\ S^z_{m}S^z_{m+1}\Bigr]\Biggr)\,,
\end{eqnarray}
with $L$ being the number of rungs.
Again within first order perturbation theory, this result can be
formulated as
% =\frac{1}{Z}\exp\left(-\beta{\hat {\cal H}}_{\rm ent}\right)
% $Z={\rm Tr}\exp(-\beta{\cal H}_{\rm ent}^{(1)})$
\begin{equation}
\rho_{\rm red}^{(1)}=\frac{1}{Z}\exp\left(-{\cal H}_{\rm ent}^{(1)}\right)
\end{equation}
with $Z={\rm Tr}\exp(-{\cal H}_{\rm ent}^{(1)})$ being a partition function.
The entanglement Hamiltonian\cite{Li08} ${\cal H}_{\rm ent}^{(1)}$  is given as:
\begin{eqnarray}
{\cal H}_{\rm ent}^{(1)}&=&\frac{4}{1+\Delta} \times \frac{\Jl}{\Jr}\nonumber\\
&& \sum_m\Bigl[\frac{1}{2}(S^+_{m} S^-_{m+1} + h.c.)
+\tilde\Delta S^z_{m}S^z_{m+1}\Bigr]\ ,
%&=:& \beta \sum_m\Bigl[\frac{1}{2}(S^+_{m} S^-_{m+1} + h.c.)
%+\tilde\Delta S^z_{m}S^z_{m+1}\Bigr]
\end{eqnarray}
and is of  nearest neighbor XXZ form, with a renormalized anisotropy parameter,
\begin{equation}
\tilde\Delta=\frac{1}{2}\left(\Delta+\Delta^2\right) \,.
\end{equation}
Note that $\Delta =0$ (XX case) and $\Delta=1$ (Heisenberg interactions) are
invariant, i.e.~the entanglement Hamiltonian is proportional to the physical 
Hamiltonian restricted to the block, as observed numerically for $\Delta=1$ earlier 
in Ref.~\onlinecite{Poilblanc10}. Since in these two specific cases the physical Hamiltonian 
on a chain and the entanglement Hamiltonian are simply proportional to each other
(to first order), one can define an ad hoc inverse temperature 
\begin{equation}
\beta=\frac{4}{1+\Delta} \times \frac{\Jl}{\Jr},
\end{equation}
such that ${\cal H}_{\rm ent}^{(1)}=\beta\ {\hat {\cal H}_\mathrm{XXZ}}$, with a 
$\Jl/\Jr$ dependence solely in $\beta$.

%\begin{equation}
%\beta=\frac{4}{1+\Delta} \times \frac{\Jl}{\Jr} 
%\end{equation}
%playing the formal role of an inverse temperature, and the entanglement 
%Hamiltonian
%\begin{equation}
%{\cal H}_{\rm ent}=\sum_m\Bigl[\frac{1}{2}(S^+_{m} S^-_{m+1} + h.c.)
%+\tilde\Delta S^z_{m}S^z_{m+1}\Bigr]
%\end{equation}

\subsection{Second order}

The second-order contribution to the ground state is somewhat lengthy and
given in appendix \ref{2nd}. For general anisotropy the result does not
seem to be amenable to a simple interpretation. In the isotropic case
$\Delta=1$, however, one finds  up to second order
\begin{eqnarray}
& & \rho_{\rm red}^{(2)}=\frac{1}{2^L}\Biggl(1-\frac{2\Jl}{\Jr}
\sum_m\vec S_{m}\vec S_{m+1}\nonumber\\
& & \qquad+\frac{1}{2}\left(\frac{2\Jl}{\Jr}\right)^2\Biggl(
\left[\sum_m\vec S_{m}\vec S_{m+1}\right]^2-\frac{3}{16}L\nonumber\\
& & \quad\qquad+\frac{1}{4}
\sum_m\left[\vec S_m\vec S_{m+2}-\vec S_m\vec S_{m+1}\right]
\Biggr)\Biggr)\,,
\label{2ndrho1}
\end{eqnarray}
which can, within the same order, be reformulated as
\begin{equation}
\rho_{\rm red}^{(2)}=\frac{1}{Z}
%\exp\left(-\beta{\cal H}_{\rm ent}^{(2)}\right)
\exp\left(-{\cal H}_{\rm ent}^{(2)}\right)
\label{2ndrho2}
\end{equation}
with %$\beta=2\Jl/\Jr$, %$Z={\rm tr}\exp(-\beta{\cal H}_{\rm ent}^{(2)})$
$Z={\rm tr}\exp(-{\cal H}_{\rm ent}^{(2)})$
and 
\begin{eqnarray}
\label{eqn:2ndEH}
{\cal H}_{\rm ent}^{(2)}&=& 2\frac{\Jl}{\Jr} \sum_m\vec S_m\vec S_{m+1}\\
&&+\frac{1}{2}\left(\frac{\Jl}{\Jr}\right)^2
\sum_m\left[\vec S_m\vec S_{m+1}-\vec S_m\vec S_{m+2}\right]\,.\nonumber
\end{eqnarray}
Thus, also within second order $\Jl/\Jr$ the entanglement Hamiltonian is quite similar
to the physical Hamiltonian on a single chain. The second order correction contains a
renormalization of the amplitude of the nearest neighbor coupling and the 
appearance of an unfrustrated ferromagnetic next nearest neighbor Heisenberg 
coupling. The latter observation is in accordance with the numerical findings in 
Refs.~\onlinecite{Cirac11,Peschel11}. We expect this picture to hold also
at higher order. According to general linked cluster ideas, sites at distance $r$ will only
be able to couple starting at order $r$ with a leading amplitude proportional to $\left(\frac{\Jl}{\Jr}\right)^r$.
Given the Heisenberg nature of the leg interactions, to leading order a Heisenberg 
interaction $\sum_m\vec S_m\vec S_{m+r}$ is the only symmetry-allowed interaction.

\begin{figure}
\centerline{\includegraphics[width=\linewidth]{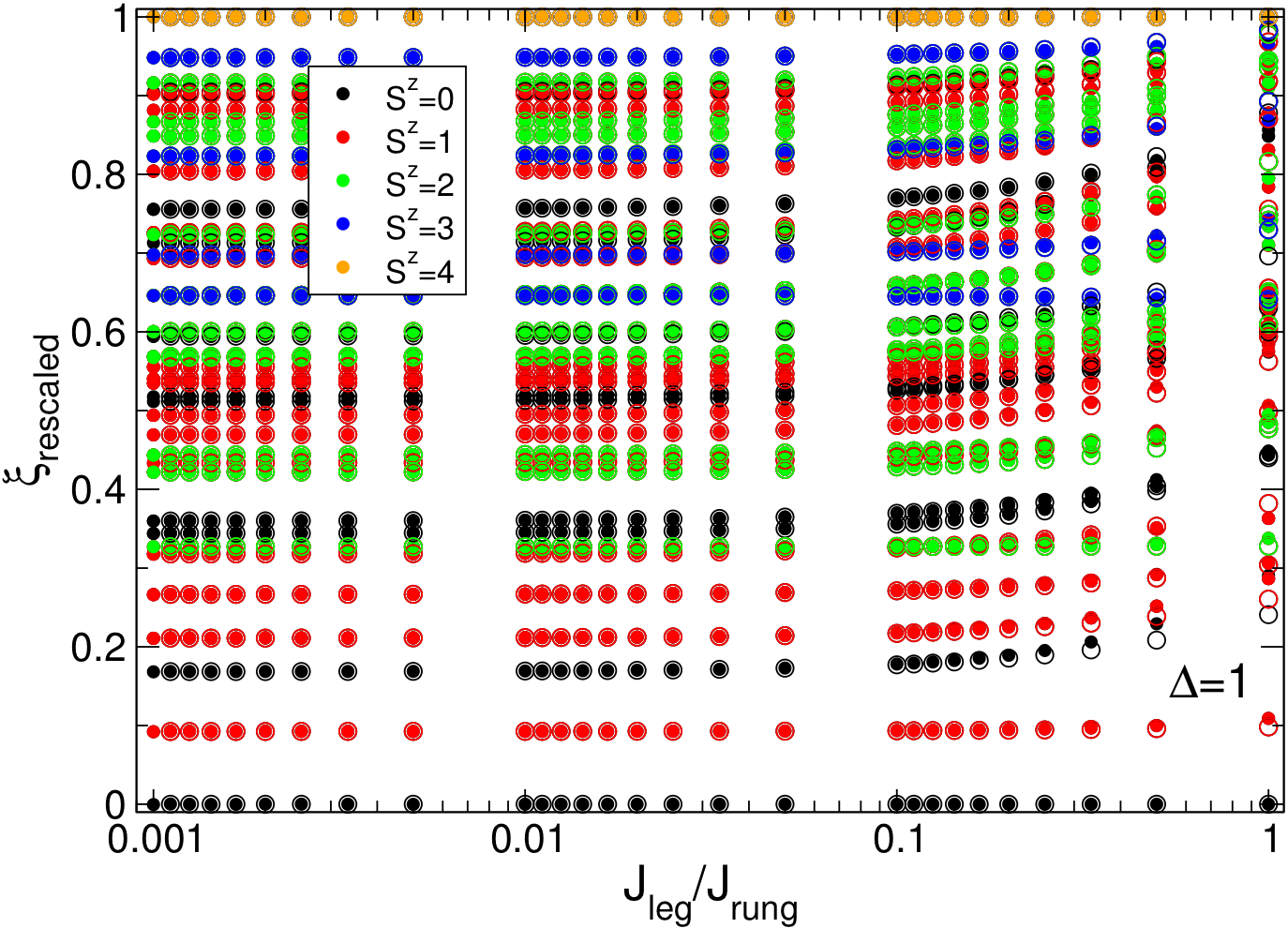}}
\caption{(Color online) Filled circles: numerical entanglement spectra obtained for a $L=8$ system. For each
$\Jl/\Jr$ value the entanglement spectrum has been rescaled to lie in the interval $[0,1]$. Empty circles: rescaled 
energy spectrum of the analytical second order entanglement Hamiltonian Eq.~(\ref{eqn:2ndEH}).}
\label{fig:secondorder_numerics}
\end{figure}

In Fig.~\ref{fig:secondorder_numerics} we present a comparison between the 
rescaled numerical entanglement spectra for $L=8$ and the perturbative result
in the form of the energy spectrum of Eq.~(\ref{eqn:2ndEH}) for $\Jl/\Jr\le1$.
The second order corrections only seem to become visible for this system size
for $\Jl/\Jr\gtrsim 0.1$, and the initial deviations from a straight line in the numerics
are correctly described by the second order corrections. However for larger
$\Jl/\Jr$ ratios further neighbor Heisenberg couplings and multi-spin interactions
become non-negligible, as observed in Ref.~\onlinecite{Cirac11}.

%Thus, also within second order in $\Jl/\Jr$ the reduced
%density matrix is essentially the canonical density matrix generated by
%the Heisenberg chain Hamiltonian of the remaining leg, up to an additional
%contribution being of higher order in $\Jl/\Jr$ and involving couplings 
%of longer range between next-nearest neighbors. 

%%%%%%%%%%%%%%%%%
%%%%%%%%%%%%%%%%%

\section{Limit of weakly coupled chains}

We now discuss the opposite limit of weakly coupled chains.  At the starting
point  $\Jr/\Jl=0$ the ground state is the product of the ground states of  the 
individual chains.

\begin{equation}
|\psi_0\rangle = |0\rangle_A|0\rangle_B
\end{equation}

In this limit the entanglement spectrum is trivial, composed of only a single 
value $\xi=0$,

To first order the wave function reads: 
\begin{equation}
|\psi_1\rangle=[1 - R Q_0{\cal H}_0]|\psi_0\rangle, 
\end{equation}
where ${\cal H}_0$ is the Hamiltonian coupling the two chains 
[cf. Eq.~(\ref{eqn:Hamiltonian})], $Q_0$ is the projector onto the subspace 
orthogonal to $|\psi_0\rangle$,
and $R$ denotes the resolvent operator $R= ({\cal H}_1-E_0)^{-1}$. \footnote{
For a finite system the perturbation theory is expected to have a finite radius of
convergence.}

After a straightforward calculation one obtains the following expression for the
first order wavefunction $|\psi_1\rangle$ in the product basis of chain eigenfunctions:
\begin{eqnarray}
&&|\psi_1\rangle = |0\rangle_A|0\rangle_B \\
&&+ \frac{\Delta\ \Jr}{\Jl} \sum_k \sum_{n,n'\neq 0} \frac{\langle n | S^z_k |0\rangle_A \langle n' | S^{z}_{-k} |0\rangle_B}{\Delta_n + \Delta_{n'}} |n\rangle_A |n'\rangle_B\nonumber\\
&&+ \frac{\Jr}{\Jl} \sum_{k;\alpha\in\{x,y\}} \sum_{n,n'\neq 0} \frac{\langle n | S^\alpha_k |0\rangle_A \langle n' | S^{\alpha}_{-k} |0\rangle_B}{\Delta_n + \Delta_{n'}}|n\rangle_A |n'\rangle_B\nonumber\\
&&= \sum_{n,n'} [\psi_1]_{n,n'} |n\rangle_A |n'\rangle_B,
\end{eqnarray}
where $k$ runs over the lattice momenta of a single chain, $\alpha$ runs over 
the spin components $\{x,y\}$,  $n$ and $n'$ label the eigenfunctions of the 
two isolated chains,
while $\Delta_{n(n')}$ denotes the single chain excitation energies 
$E_{n(n')}-E_0$. One recognizes that the nontrivial part of the wave function 
is composed of
contributions which also enter the Lehmann representation 
of the ${\cal S}^{\alpha\alpha}(k,\omega)$ spectral functions of a single chain,
\begin{eqnarray}
{\cal S}^{\alpha}(k,\omega)&=& 
\sum_n |c^{k,\alpha}_n|^2 \times \delta(\omega-\Delta_n)\,,\\
c^{k,\alpha}_n &=& \langle n | S^\alpha_k | 0 \rangle\,,
\end{eqnarray}
where the matrix elements $c^{k,\alpha}_n$ enter as
\begin{eqnarray}
[\psi_1]_{n,n'} &=& \delta_{n,0} \delta_{n',0}\nonumber\\
&+& \frac{\Delta \Jr}{\Jl} \sum_{k}  \frac{c^{k,z}_{n} c^{k,z}_{n'}}{\Delta_n+\Delta_{n'}}\nonumber\\
&+&\frac{1}{2}\frac{\Jr}{\Jl} \sum_{k}  \frac{c^{k,+}_{n} c^{-k,-}_{n'}}{\Delta_n+\Delta_{n'}}\nonumber\\
&+& \frac{1}{2}\frac{\Jr}{\Jl} \sum_{k}  \frac{c^{k,-}_{n} c^{-k,+}_{n'}}{\Delta_n+\Delta_{n'}}.
\label{eqn:spectral_es}
\end{eqnarray}

Given the wave function in this form, the entanglement spectrum can simply be obtained by a singular value decomposition
of the matrix $[\psi_1]_{n,n'}$, which amounts to finding the Schmidt decomposition in this bipartition.
The singular values are thus the Schmidt values, and when squared they correspond to the
eigenvalues of the reduced density matrix in the same bipartion setup. Due to the translational invariance along the chains 
as well as the total $S^z$ preserving form of the XXZ Hamiltonians, the wave function matrix exhibits a block diagonal structure 
in $k$ and only $n$ and $n'$ sectors with total $S^z=0,\pm 1$ appear. In the specific case of Heisenberg Hamiltonians $(\Delta=1)$
one can then infer by virtue of the Wigner-Eckart theorem that only total spin $S=1$ entanglement levels appear above the singlet 
ground state to leading order. In addition the singlet ground state leads to vanishing spectral weight at $k=0$ and therefore the
absence of $S=1$ entanglement levels at $k=0$ to the expansion order considered. 

%%%%%%%%%%%%%%%%%
\begin{figure}
\centerline{\includegraphics[width=\linewidth]{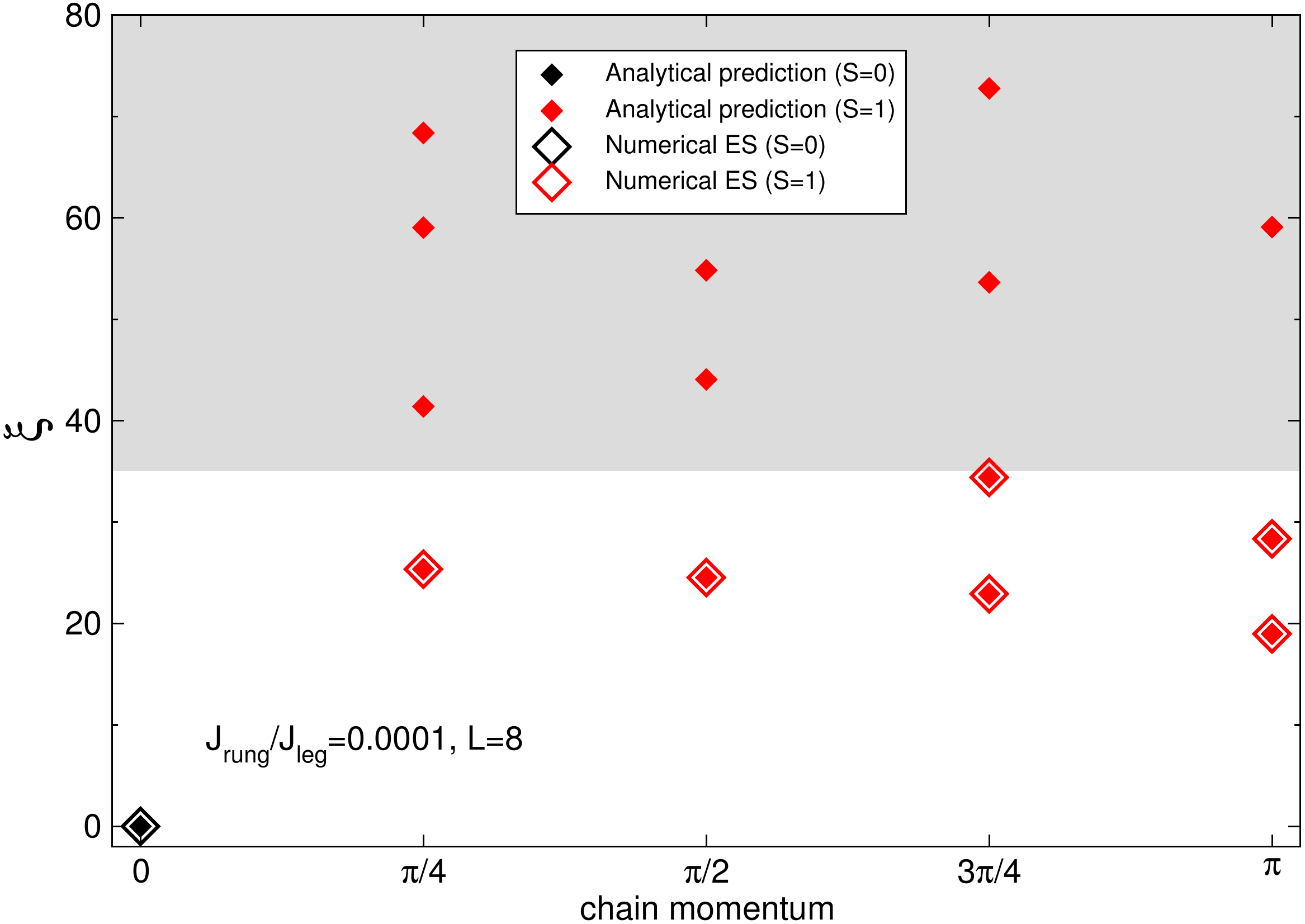}}
\caption{(Color online) Filled diamonds: entanglement spectrum prediction for a $L=8$ Heisenberg spin ladder, 
based on the first order perturbation theory in the coupling between the two chains [Eqn.~(\ref{eqn:spectral_es})]. 
Empty diamonds:  $S=0$ and $S=1$ entanglement levels obtained from exact diagonalization at $\Jr/\Jl=0.0001$, 
cf. upper panel of Fig.~\ref{fig:es_D1}. The grey region denotes the region where the numerical entanglement spectrum
is affected by the finite precision arithmetics and where no numerical data is shown.}
\label{fig:weaklycoupledchains}
\end{figure}
%%%%%%%%%%%%%%%%%

In Fig.~\ref{fig:weaklycoupledchains} we show a comparison between the numerical entanglement spectrum for $L=8$ at
$\Jr/\Jl=0.0001$ and the prediction based on Eqn.~(\ref{eqn:spectral_es}). The filled symbols represent {\em all} the predicted
entanglement levels appearing at this order. The numerical results match perfectly the analytical prediction up to the finite precision threshold 
of the numerics.

\section{Discussion and Conclusion}

In summary, we have investigated the entanglement spectra of spin-$1/2$ XXZ ladders using both 
exact numerical diagonalizations and perturbation theory approaches. In the limit 
of strong rung coupling perturbation theory predicts in leading order an entanglement 
Hamiltonian being also of nearest-neighbor XXZ form with a renormalized anisotropy
parameter. For XX and isotropic Heisenberg ladders no such renormalization takes place
and in this case one can define an ad hoc effective temperature proportional to the $\Jr/\Jl$
ratio. Moreover, in next-to-leading order the entanglement Hamiltonian exhibits
spin couplings of longer range, and all the above findings are in perfect
agreement with our numerical exact diagonalization results. The first order perturbation theory 
result has also been obtained in parallel work by Peschel and Chung~\cite{Peschel11}. Similar numerical observations have already made for quantum Hall
bilayers at filling factor $\nu=1$\cite{Schliemann11}. 
It remains an open task to
devise  analogous perturbational arguments for such systems with
long-ranged interactions .

In the opposite regime of strong coupling along the legs, we point out an 
interesting connection of the entanglement spectrum to the Lehmann 
representation of single chain spectral functions of the operators contained in the physical 
Hamiltonian coupling the two chains. This aspect also holds true for other ladder systems,
such as e.g. Hubbard ladders, where the entanglement spectrum will be determined by elements
of the single particle addition and removal spectral functions in the weakly coupled chain regime.

\acknowledgments
AML thanks V. Alba and M. Haque for collaboration on related topics.
The work of JS was supported by DFG via SFB631.

\appendix

\section{Second-order correction to the ground state and the 
reduced density matrix}
\label{2nd}
The second-order correction to the ground state can be obtained as
\begin{widetext}
\begin{eqnarray}
| 2\rangle & = & \left(\frac{\Jl}{\Jr}\right)^2\sum_{|m-n| >1}\Biggl[
\frac{1}{4(1+\Delta)^2}
\left(| t^+_m,t^-_{m+1},t^+_n,t^-_{n+1}\rangle+
| t^+_m,t^-_{m+1},t^-_n,t^+_{n+1}\rangle\right)\nonumber\\
& & -\frac{\Delta}{4(3+\Delta)}\left(\frac{1}{2}+\frac{1}{1+\Delta}\right)
\left(| t^+_m,t^-_{m+1},t^0_n,t^0_{n+1}\rangle+
| t^-_m,t^+_{m+1},t^0_n,t^0_{n+1}\rangle\right)
+\frac{\Delta^2}{32}| t^0_m,t^0_{m+1},t^0_n,t^0_{n+1}\rangle\Biggr]\nonumber\\
& + & \left(\frac{\Jl}{\Jr}\right)^2\sum_m\Biggl[
-\frac{1}{2(1+\Delta)^2}
\left(| t^+_m,t^-_{m+2}\rangle+| t^-_m,t^+_{m+2}\rangle\right)
+\frac{\Delta^2}{8}| t^0_m,t^0_{m+2}\rangle\Biggr]\nonumber\\
& + & \left(\frac{\Jl}{\Jr}\right)^2\sum_m\Biggl[
\frac{\Delta}{8(1+\Delta)}\left(\frac{1}{2}+\frac{1}{1+\Delta}\right)
\left(| t^+_m,t^-_{m+1}\rangle+| t^-_m,t^+_{m+1}\rangle\right)
-\frac{1}{4(1+\Delta)}| t^0_m,t^0_{m+1}\rangle\Biggr]\nonumber\\
& - & \left(\frac{\Jl}{\Jr}\right)^2\frac{L}{8}
\left(\frac{2}{(1+\Delta)^2}+\frac{\Delta^2}{4}\right)| 0\rangle\,.
\end{eqnarray}
\end{widetext}
Note that no terms with an odd number of triplets occur. In the isotropic
Heisenberg case $\Delta=1$, the second-order
contribution to the reduced density matrix is the sum of the following
expressions, 
\begin{eqnarray}
& & {\rm tr}_{\rm 1leg}\left(| 0\rangle\langle 2| +|2\rangle\langle 0| \right)
\nonumber\\
& & =\left(\frac{\Jl}{\Jr}\right)^2\frac1{2^L}\Biggl(
\sum_{\mid m-n\mid >1}\left(\vec S_m\vec S_{m+1}\right)\left(\vec S_n\vec S_{n+1}\right)
\nonumber\\
& & \quad+\sum_m\left[\vec S_m\vec S_{m+2}-\vec S_m\vec S_{m+1}\right]
-\frac{3}{16}L\Biggr)\,,\\
& & {\rm tr}_{\rm 1leg}\left(| 1\rangle\langle 1| \right)
\nonumber\\
& & =\left(\frac{\Jl}{\Jr}\right)^2\frac1{2^L}\Biggl(
\sum_{\mid m-n\mid >1}\left(\vec S_m\vec S_{m+1}\right)\left(\vec S_n\vec S_{n+1}\right)
\nonumber\\
& & \quad+\frac{1}{2}
\sum_m\left[\vec S_m\vec S_{m+2}-\vec S_m\vec S_{m+1}\right]
+\frac{3}{16}L\Biggr)\,.
\end{eqnarray}
Now using the identities
\begin{equation}
\frac{1}{2}\vec S_1\vec S_3=
\left(\vec S_1\vec S_2\right)\left(\vec S_2\vec S_3\right)
+\left(\vec S_2\vec S_3\right)\left(\vec S_1\vec S_2\right)
\end{equation}
and
\begin{equation}
\frac{3}{16}-\frac{1}{2}\vec S_1\vec S_2=\left(\vec S_1\vec S_2\right)^2
\end{equation}
for spin-$1/2$-operators, one derives the result (\ref{2ndrho1}).

\end{document}